\RequirePackage{fix-cm}
\documentclass[twocolumn]{svjour3}          
\smartqed  
\usepackage[dvipdfmx]{graphicx}

\usepackage[misc]{ifsym}
\usepackage{subfigure}
\usepackage{amsmath,txfonts}

\usepackage[dvipdfmx]{color} 

\usepackage{soul}

\begin{document}

\title{Two-step pulse observation to improve resonance contrast for coherent population trapping atomic clock
}


\author{Yuichiro Yano\and Shigeyoshi Goka \and Masatoshi Kajita 
}


\institute{Yuichiro Yano (\Letter) \and Shigeyoshi Goka\at
            Department of Electrical \& Engineering, Tokyo Metropolitan University \\
			1-1 Minami-Osawa, Hachioji-shi, Tokyo, Japan 192-0397\\
              Present E-mail address: y-yano@nict.go.jp           
           \and
           Masatoshi Kajita\at
           National Institute of Information and Communications Technology (NICT)
           }

\date{Received: date / Accepted: date}

\maketitle

\begin{abstract}
We study resonance contrast by a two-step pulse observation method to enhance the frequency stability of coherent population trapping (CPT) atomic clocks.
The proposed method is a two-step Raman--Ramsey scheme with low intensity during resonance observation and high intensity after the observation.
This method reduces the frequency variation in the light intensity and maintains a high signal-to-noise ratio. 
The resonance characteristics were calculated by density matrix analysis of a $\Lambda$-type three-level system that was modeled on the $^{133}$Cs D1 line, and the characteristics were also measured using a vertical-cavity surface-emitting laser and a Cs vapor cell.

 \PACS{PACS code1 \and PACS code2 \and more}
\end{abstract}

\section{Introduction}
\label{intro}

\begin{figure}[t]
\includegraphics[width=\hsize,keepaspectratio=true]{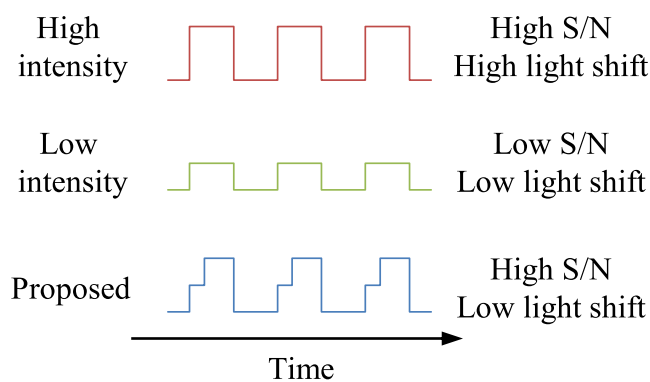}
\caption{Proposed Ramsey pulse sequence.}
\label{fig:scheme}
\end{figure}

Coherent population trapping (CPT) atomic clocks are in great demand for many applications, such as telecommunications,
navigation systems, and the synchronization of networks\cite{vig1992military}, and such clocks are required for their high frequency stability.

The frequency stability in atomic clocks can be classified into two types based on the averaging time of the Allan deviation, which are the short-term and long-term frequency stability.
The short-term stability is determined as the inverse of the product of the signal-to-noise ratio (SNR) and the Q value\cite{vanier1989quantum}.
The SNR of laser-pumped vapor cell atomic clocks is often limited by a combination of light source AM noise and FM-AM conversion noise in the atomic absorption.
Therefore, the contrast, which is given by the resonance signal amplitude divided by the background signal level, is used as the SNR of the CPT resonance.
The Q value is given by clock transition frequency divided by the resonance linewidth.
The Q value is degenerated by the power broadening effect.
Therefore, it is necessary to improve the resonance contrast and reduce power broadening to enhance the short-term frequency stability.
The long-term frequency stability is mainly limited by the light shift (the ac Stark shift induced by the laser light) because it changes with the light power fluctuation or the aging of optical elements.

The Raman--Ramsey (RR) scheme for enhancing the frequency stability of CPT atomic clocks has been investigated by a number of researchers\cite{zanon2005high,esnault2013cold,xi2010coherent,yoshida2013line,yano2014theoretical}.
This scheme simultaneously reduces line broadening and the light shift \cite{yoshida2013line,yano2014theoretical,castagna2009investigations}.
Recently, a short-term stability of 3.2$\times 10^{-13}\tau^{-1/2}$ has been obtained, and stability as low as 3$\times10^{-14}$ has been achieved at an averaging time of 200 s using a Cs vapor cell with the RR scheme\cite{danet2014dick}. 
Also, when cold Rb atoms were used, a short-term frequency stability of 4$\times 10^{-11}\tau^{-1/2}$ was obtained, and a long-term frequency stability of 3$\times 10^{-13}$ was achieved for an averaging time of 5 h\cite{donley2014frequency}.

In our previous paper, we investigated the light shift in the RR scheme both theoretically and experimentally with the aim of enhancing the long-term frequency stability of CPT atomic clocks\cite{yano2014theoretical}.
The results showed that the light shift in the RR scheme is lower than that under cw illumination.
We also found that the fluctuation of the clock frequency due to the light intensity fluctuation is reduced by reducing the observation time, which is defined as the time interval between the rise of the pulsed laser and the observation.

The main reason for this is that the atoms evolve towards a steady state during observation.
In addition, we found that the light shift in the RR scheme is a nonlinear function of the light intensity,
and reduced clock frequency variation is obtained under a high light intensity beyond a set threshold value.
To reduce the light shift, it is preferable to perform measurements with the observation time as short as possible and the light intensity slightly higher than the threshold.

A higher signal can also be obtained when setting both a short detection time and high light intensity{\cite{guerandel2007raman}}.
Certainly, the signal increases with increasing light intensity; however the noise also increases owing to the short detection time and the AM noise in the high intensity.
Thus, the contrast does not increase with increasing light intensity.
While a low light intensity leads to a weak signal, a detected signal with low laser intensity decreases more slower than one with a high intensity.
Consequently, the signal can be stronger at a low intensity for large detection times.

We proposed a two-step pulse observation method to resolve this issue (Fig. \ref{fig:scheme}).
This method is an RR scheme with low light intensity during observation\cite{yano2015two}.
We investigated the light shift by two-step pulse observation.
As a result, it was shown that the clock frequency variation in the light intensity using this method is lower than that of the conventional method owing to the reduced repumping rate toward the steady state during the observation time.

In this work, we discuss the improvement of the resonance contrast when using the two-step pulse observation method via numerical calculation and experiment.
A comparison with the RR scheme is performed by calculation using the density matrix analysis.
The characteristics were also measured using a vertical-cavity surface-emitting laser (VCSEL) and a Cs vapor cell.
We investigated the contrast and Q value as functions of the free evolution time and observation time because they are important parameters that determine the contrast and Q value.
From the results, it is shown that this method reduces the repumping to the steady state and provides a larger contrast than the conventional method.


\section{Theory}

\subsection{Two-step pulse observation}

\begin{figure}[t]
\includegraphics[width=\hsize,keepaspectratio=true]{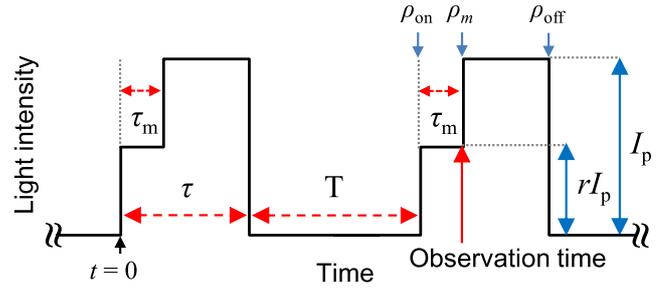}
\caption{Ramsey pulse sequence: $\tau $ is the excitation time, $T$ is the free evolution time, and $\tau_m$ is the resonance signal observation time. 
}
\label{fig:sequence}
\end{figure}

\begin{figure}[t]
\includegraphics[width=\hsize,keepaspectratio=true]{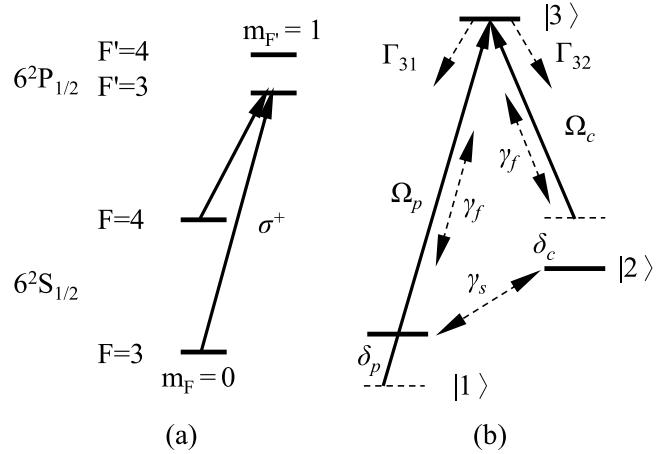}
\caption{(a) Excitation scheme with the $\sigma^+$ light field on the $D_1$ line of Cs. (b) Closed $\Lambda $-type three-level model used to calculate CPT phenomenon: $\delta_p$ and $\delta _c$ are the detunings of the probe laser and coupling laser, respectively.
$\Omega_p$ and $\Omega_{c}$ are Rabi frequencies.
$\Gamma_{31}$ and $\Gamma _{32}$ are the relaxation terms between an excited state and the two ground states.
$\gamma_f$ is the decoherence rate between the excited state and the ground states, and $\gamma_{s}$ is the decoherence rate between the two ground states.
}
\label{fig:model}
\end{figure}
The observation scheme for the two-step pulse observation method is shown in Fig. \ref{fig:sequence}.
$I_p$ is the total light intensity irradiating to the gas cell.
The Ramsey--CPT resonance is observed by measuring the transmitted light intensity at observation time $\tau _m$ after the pulse rise.
The two-step pulse observation maintains a low light intensity of $rI_p$ until the observation point, where $r$ is defined as the observation intensity ratio, and takes a values from 0 to 1.
When $r=1$ is set, the two-step pulse observation method is treated with the same scheme as that used for the conventional RR scheme.
After the measurement is taken, the light intensity is converted into $I_p$ and the atoms are prepared for the next measurement.
A laser pulse with a duration of ($\tau-\tau_m$) irradiates the atoms for pumping of the steady dark state.
After the free evolution time $T$, a laser pulse irradiates the atoms again with light intensity $rI_p$.
The light intensity $I(t)$ for this scheme is written as
\begin{equation}
I(t)=\left\{
\begin{array}{l l}
rI_p &(0<t\leq\tau_m)\\
I_p&(\tau_m<t\leq\tau)\\
0&(\tau<t\leq\tau+T)\\
\end{array}
\right..
\label{eq:scheme}
\end{equation}

\subsection{Theoretical model}

Figure \ref{fig:model}(a) shows the excitation scheme using the $\sigma ^{+}$  light field for the $^{133}$Cs-$D_1$ line.
In the CPT phenomenon, the two ground states of 6$^{2}$S$_{1/2}$ are coupled simultaneously to a common excited state of 6$^{2}$P$_{1/2}$.
In this system, the dynamical behavior of the density matrix $\rho $ is governed by the quantum Liouville equation,
\begin{equation}
\label{eq1}
\frac{\partial}{\partial t}\rho (t)=\frac{i}{\hbar }[\rho ,H]+R\rho,
\end{equation}
where $H$ is the Hamiltonian matrix for this three-level system and $R$ represents the relaxation terms. Using the rotating-wave approximation with the simplified $\Lambda $-type model depicted in Fig. \ref{fig:model}(b), Eq. (\ref{eq1}) can be rewritten as 

\begin{eqnarray}
\begin{split}
\dot{\rho}_{11}&=i\frac{\Omega_p}{2}(-\rho_{13}+\rho_{31})+\Gamma_{31} \rho_{33}+\gamma_s(\rho_{22}-\rho_{11}), \\
\dot{\rho}_{22}&=i\frac{\Omega_c}{2}(-\rho_{23}+\rho_{32})+\Gamma_{32} \rho_{33}-\gamma_s(\rho_{22}-\rho_{11}), \\ 
\dot{\rho}_{33}&=i\frac{\Omega_p}{2}(\rho_{13}-\rho_{31})+i\frac{\Omega_c}{2}(\rho_{23}-\rho_{32})-\Gamma_3 \rho_{33}, \\
\dot{\rho}_{12}&=i \rho_{12}(\delta_p-\delta_c)-i\frac{\Omega_c}{2}\rho_{13}+i\frac{\Omega_p}{2}\rho_{32}-\gamma_s \rho_{12}, \\
\dot{\rho}_{13}&=i\rho_{13}\delta_p-i\frac{\Omega_c}{2}\rho_{12}+i\frac{\Omega_p}{2}(\rho_{33}-\rho_{11})-\gamma_f \rho_{13}, \\
\dot{\rho}_{23}&=i\rho_{23}\delta_c-i\frac{\Omega_p}{2}\rho_{21}+i\frac{\Omega_c}{2}(\rho_{33}-\rho_{22})-\gamma_f \rho_{23}, 
\end{split}
\label{eq2}
\end{eqnarray}
\noindent where $\Omega_p$ and $\Omega_c$ are Rabi frequencies and $\delta_p$ and $\delta_c$ are frequency detunings, $\Gamma_{31}$ and $\Gamma_{32}$ are relaxation terms between an excited state and the two ground states, with the trace of the density matrix satisfying the closed-system condition
\begin{equation}
\label{eq3}
\rm{Tr}(\rho )=\rho_{11} +\rho_{22} +\rho_{33} =1.
\end{equation}

\noindent In Fig. \ref{fig:model}(b), $|1\rangle$ and $|2\rangle$ correspond to the two ground states $|F = 3, m_{F} =0\rangle$ and $| F = 4, m_{F} = 0 \rangle$ in the 6$^{2}$S$_{1/2}$ state, respectively, and $|3\rangle$ corresponds to the state in 6$^{2}$P$_{1/2}$.
We assume $\Gamma_{31}=\Gamma_{32}$($=\gamma_f$).
The ground-state relaxation rate $\gamma_s$ is a minuscule quantity, $\gamma_s \ll \gamma_{f}$.
In this calculation, the total emission rate was set at 370 MHz, which was obtained experimentally from the absorption lines of a Cs cell with Ne buffer gas at a pressure of 4.0 kPa.
The obtained emission rate of 370 MHz is consistent with the value estimated in Ref. \cite{pitz2009pressure}.
For bichromatic light, we have the relation $\delta_p=-\delta_c=\Delta_0/2$.

The numerical calculation method for the density matrix $\rho$ is described in the appendix A.
The density matrix $\rho$ is used to obtain the absorption index, and the resonance contrast can be derived as
\begin{equation}
{\rm Contrast (\%)}=k \frac{\hbar \omega n_{atom} L}{3rI_p}\Delta\Omega{\rm Im}(\rho)\cdot100,
\label{eq:contrast}
\end{equation}
where $k$ is a proportional constant that is determined by the sum of the laser intensities not interacting with alkali atoms such as higher-order sidebands, $\hbar$ is Planck's constant, $\omega$ is the angular frequency of the laser, $c$ is the speed of light, $n_{atom}$ is the number of atoms per unit volume, $L$ is the optical length of the gas cell, and $\Delta\Omega{\rm Im}(\rho)$ is the difference between the maximum and minimum values of $\Omega_{13}{\rm Im}(\rho_{13})+\Omega_{23}{\rm Im}(\rho_{23})$.

\subsection{Calculated Ramsey--CPT resonance spectrum with two-step pulse observation}
\begin{figure}[t]
\includegraphics[width=\hsize,keepaspectratio=true]{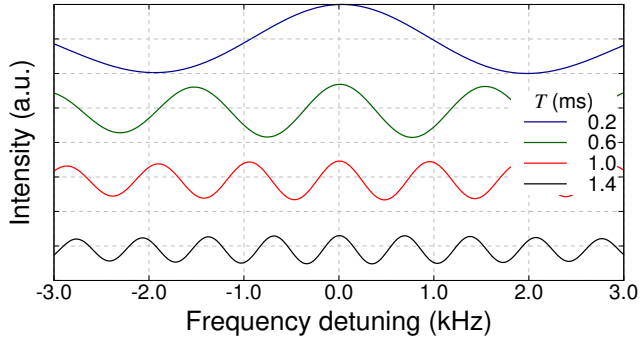}
\caption{(Color online) Calculated resonance spectrum for different free evolution times $T$: the excitation time $\tau$ is 1.0 ms, the observation time $\tau_m$ is 10 $\mu$s, the observation intensity ratio $r$ is 1.0.}
\label{fig:free_spectrum}
\end{figure}

The calculated resonance spectra for different free evolution times $T$ are shown in Fig. {\ref{fig:free_spectrum}}.
The excitation time $\tau$ and observation time $T$ were 1.0 ms and 10 $\mu$s, respectively.
The density matrix was calculated using Eq. ({\ref{eq:eigen}}) in the appendix A.
The Ramsey--CPT amplitude decreased with increasing free evolution time because of the coherence relaxation between ground states.
The resonance linewidth decreased with increasing $T$ since the Ramsey scheme was used\cite{boudot2009current}.

\begin{figure}[t]
\begin{minipage}{1\hsize}
s\subfigure[$r=1.00$]{\includegraphics*[width=\hsize,keepaspectratio=true]{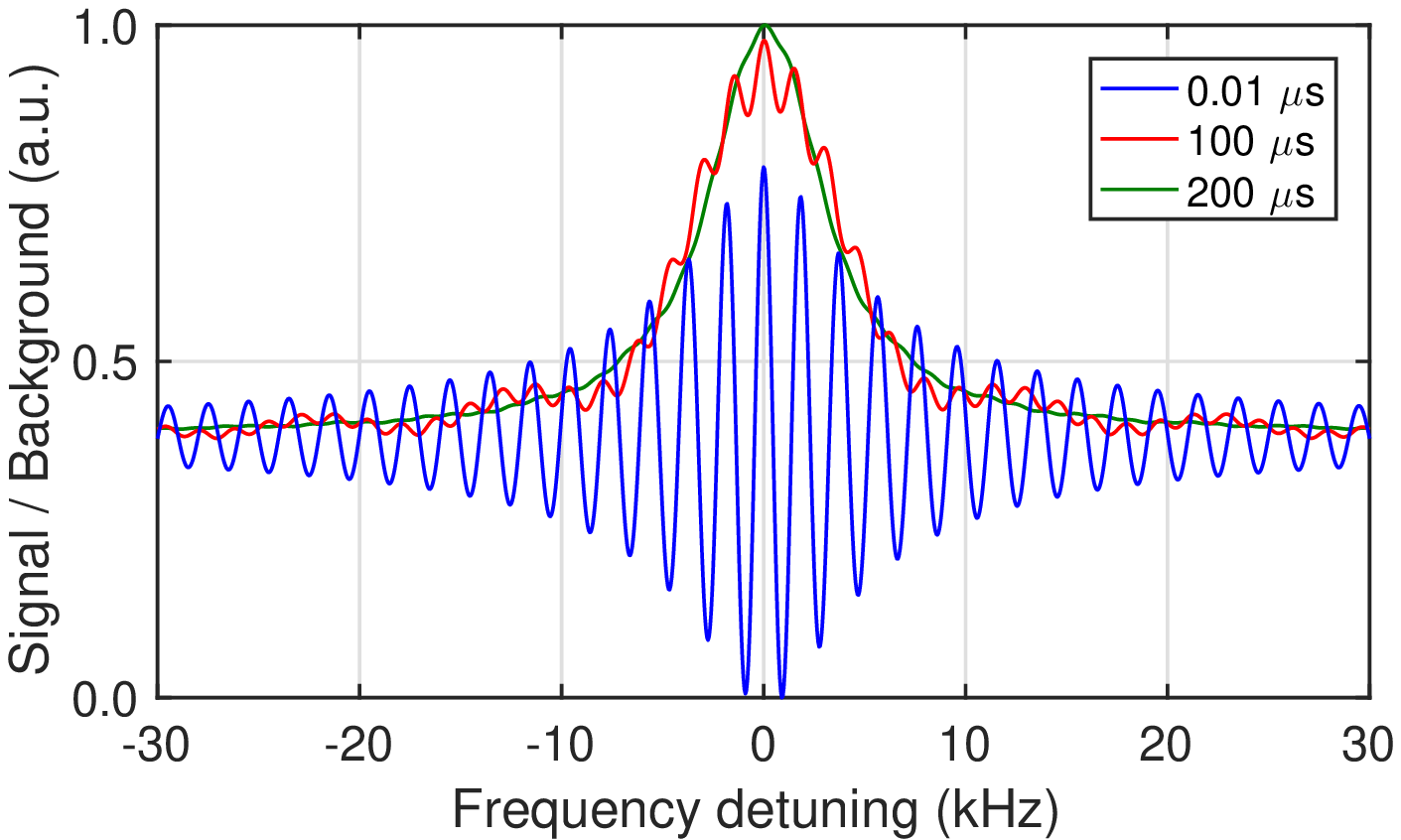}}
\subfigure[$r=0.25$]{\includegraphics*[width=\hsize,keepaspectratio=true]{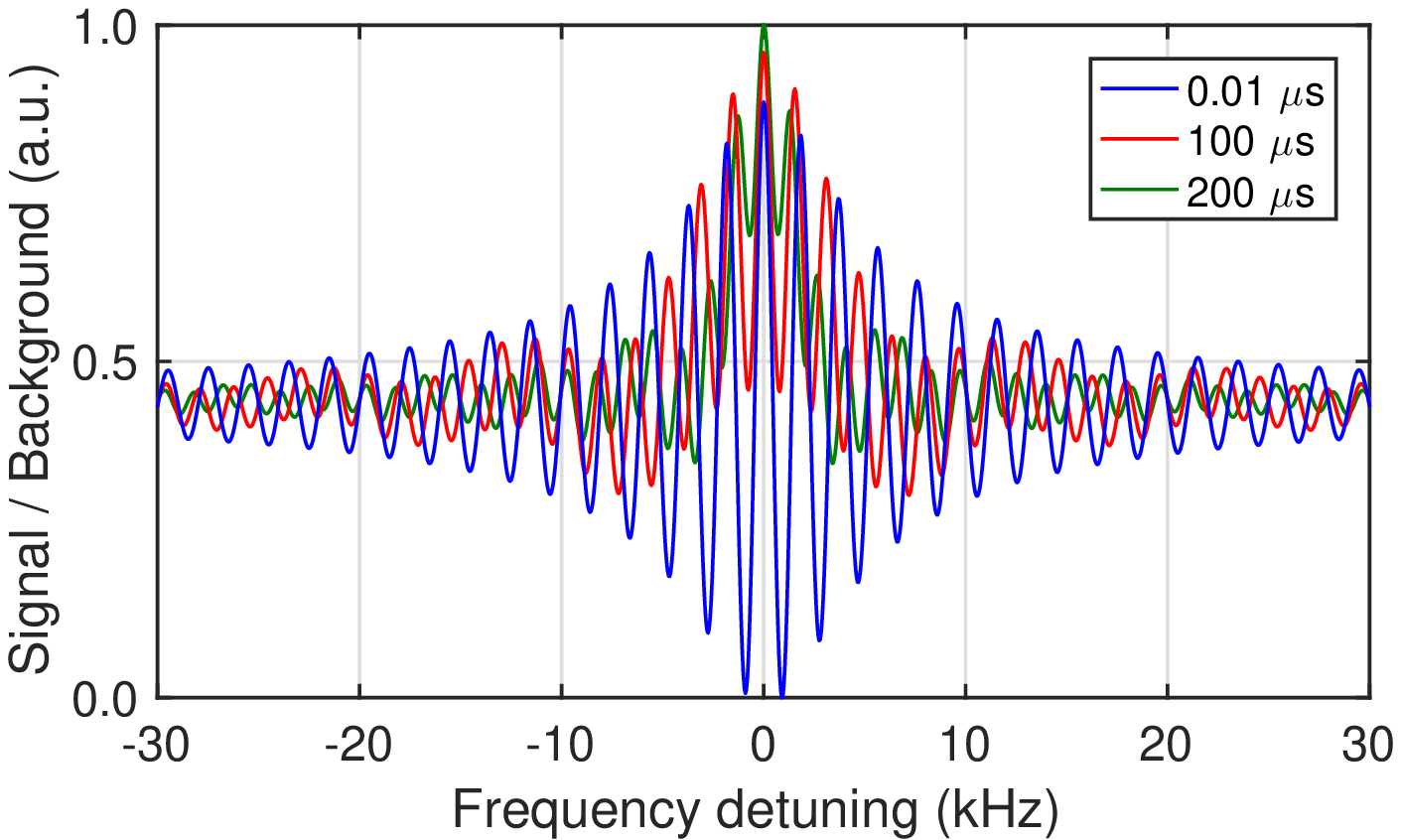}}0
\subfigure[$r\sim0.00$]{\includegraphics*[width=\hsize,keepaspectratio=true]{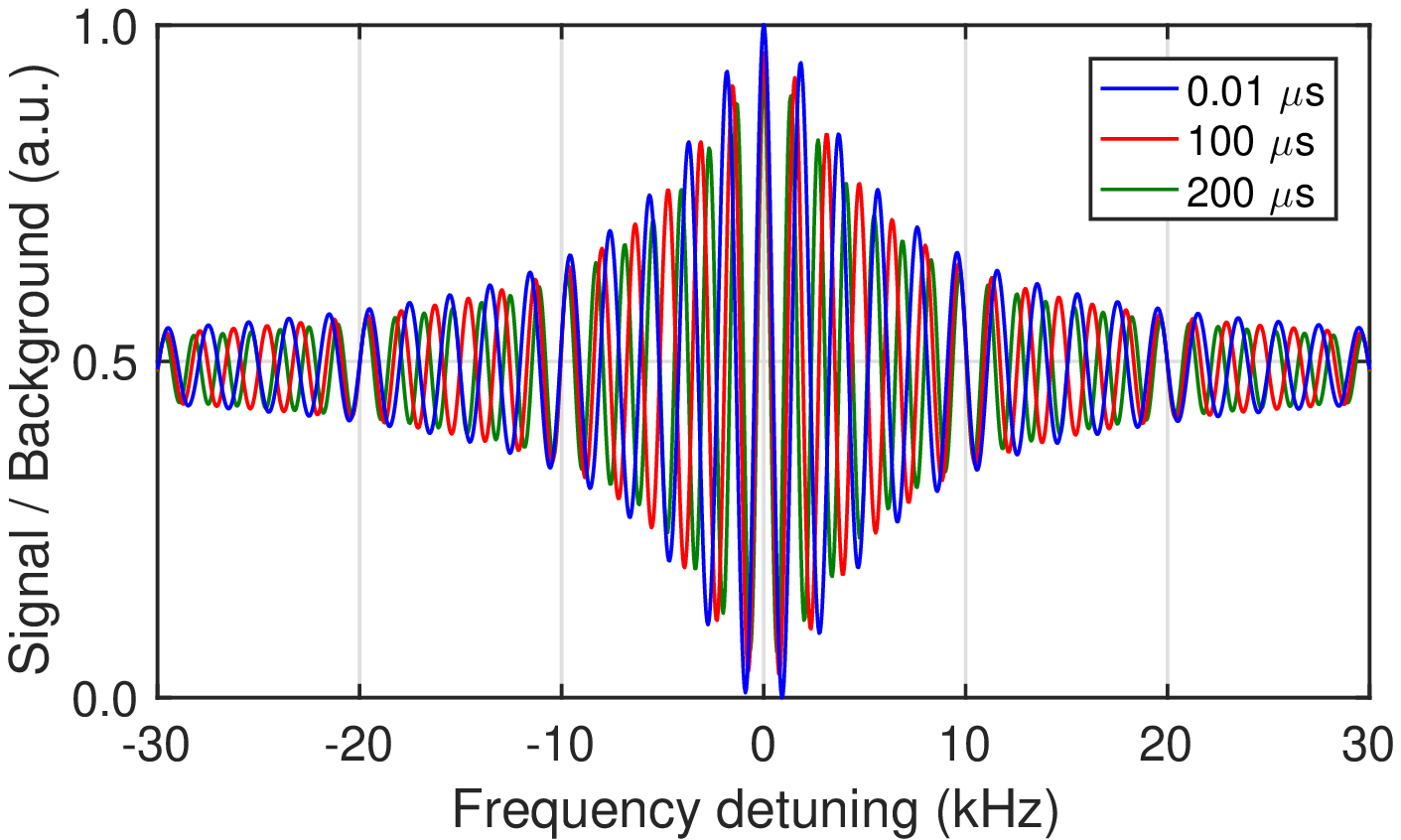}}
\end{minipage}
\caption{(Color online) Calculated resonance spectrum for different observation time $\tau_m$ and $r$: the free evolution time $T$ is 500 $\mu$s, the excitation time $\tau$ is 1.0 ms, the observation intensity ratio $r$ is 1.0.}
\label{fig:observation_spectrum}
\end{figure}

Figure {\ref{fig:observation_spectrum}} shows the calculated resonance spectra for different observation times $\tau_m$ and observation intensity ratios $r$.
$T$ and $\tau$ are 0.5 and 1.0 ms, respectively.
The total light intensity $I_p$ is 2.4 mW/cm$^2$.
The Ramsey fringe was clearly observed for a small observation time ($\tau_m \sim 0$).
However, it vanished for a large observation time.
The resonance amplitude of the Ramsey fringe strongly depended on $\tau_m$, and it substantially decreased with increasing $\tau_m$.
This is because atoms evolved toward a steady state during the time from the pulse rise to the observation point.
The resonance amplitude of the Ramsey fringe in the two-step pulse observation can be retained for a longer observation time.
Because the repumping rate is proportional to the light intensity, a smaller repumping rate is obtained when setting a smaller $r$.
Therefore, even if a long observation time is set, a large resonance amplitude can be obtained.
Since the linewidth of the center fringe is independent of the observation time, the two-step pulse observation enhances the SNR without degenerating the Q value.

\section{Experimental setup}

\begin{figure}[t]
\includegraphics[width=\hsize,keepaspectratio=true]{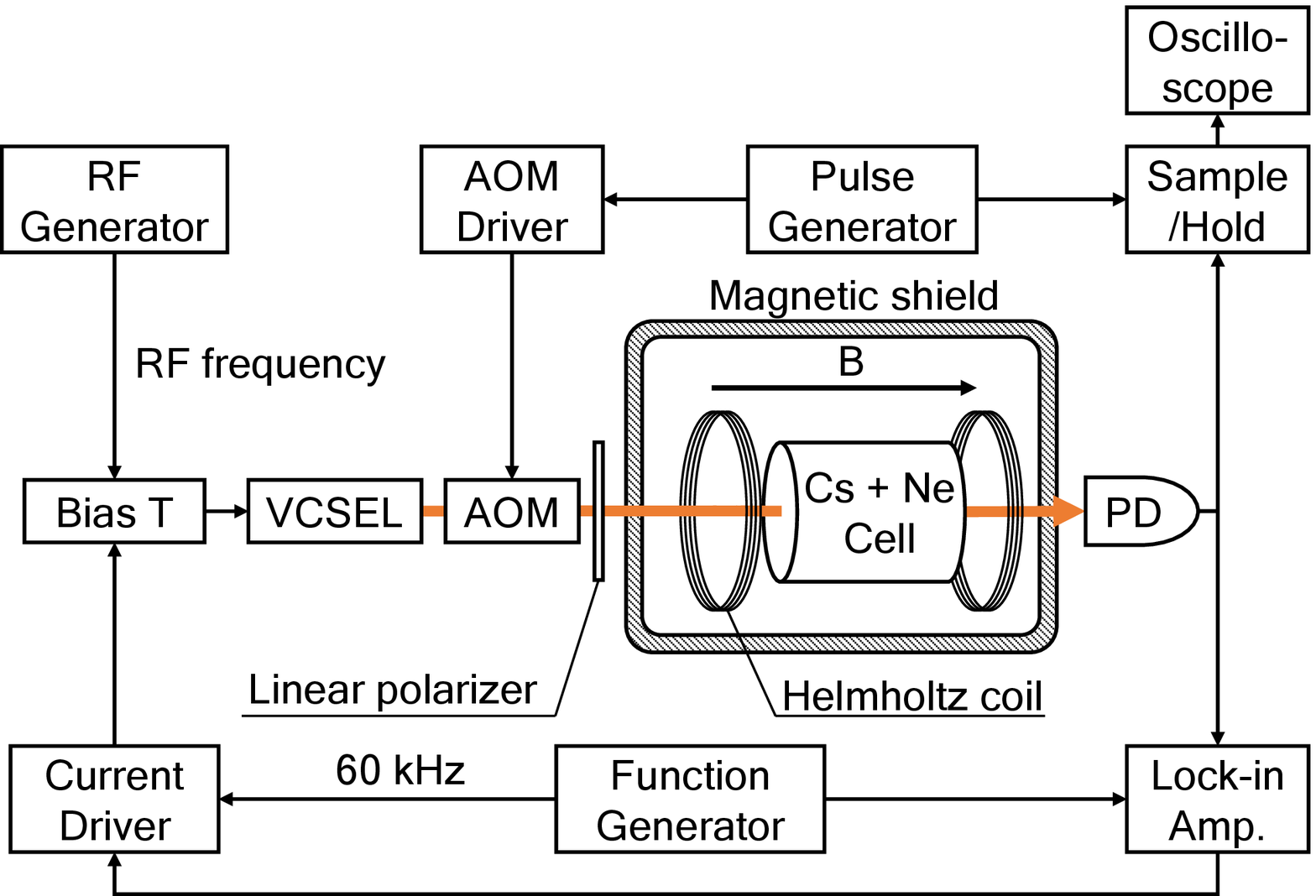}
\caption{Schematic of experimental setup. PD: photodiode. }
\label{fg:experiment}
\end{figure}

Figure \ref{fg:experiment} shows the experimental setup.
The measurement system is based on a previous Ramsey--CPT observation system.

A single-mode VCSEL (Ricoh Company Ltd., Japan) was used to simultaneously excite the two ground states to the common excited state.
The wavelength of the VCSEL used to excite $^{133}$Cs at the $D_1$ line was 895 nm.
The VCSEL was driven by a dc injection current using a bias T and was modulated at 4.6 GHz using an analog signal generator to generate first-order sidebands around the laser carrier. 

For pulse excitation, an acousto-optical modulator (AOM) was used to modulate the light intensity.
A first-order diffraction beam was irradiated to the gas cell.
The AOM had a nominal rise and fall time of 65 ns.
The total light intensity incident on the gas cell was adjusted using the control voltage of the AOM and was calibrated by an optical power meter.
A Pyrex gas cell containing a mixture of $^{133}$Cs atoms and Ne buffer gas at a pressure of 4.0 kPa was used.
This cell was cylindrical with a diameter of 20 mm and an optical length of 22.5 mm.
Its temperature was maintained at 42.0 $^\circ$C.
The gas cell and Helmholtz coil were covered with a magnetic shield to prevent external magnetic fields from affecting the magnetic field inside the cell.
The Helmholtz coil produced an internal magnetic field in the gas cell.
The axis of the 10 $\mu $T magnetic field was set parallel to the direction of the laser light ($C$-axis direction).

\section{Results}

\subsection{Contrast and Q value as functions of free evolution time $T$}

\begin{figure}[t]
\includegraphics[width=\hsize,keepaspectratio=true]{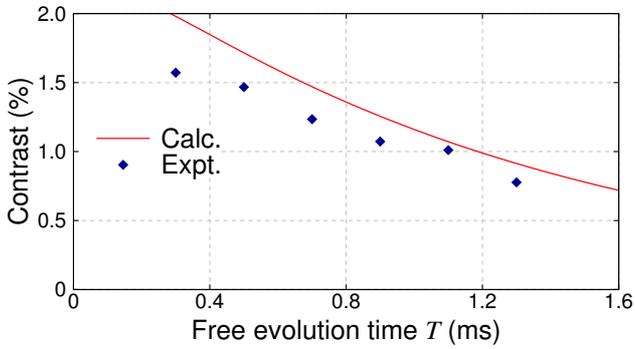}
\caption{Contrast as a function of the free evolution time $T$.}
\label{fig:free_contrast}
\end{figure}

\begin{figure}[t]
\includegraphics[width=\hsize,keepaspectratio=true]{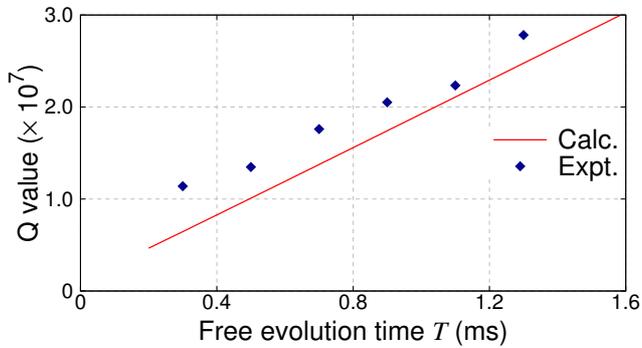}
\caption{Q value as a function of the free evolution time $T$.}
\label{fig:free_Q}
\end{figure}

The contrast as a function of the free evolution time is shown in Fig. \ref{fig:free_contrast}.
The excitation time and observation time were set to 1.0 ms and 5.0 $\mu$s in both the experiment and the calculation, respectively.
The total light intensity $I_p$ was 2.4 mW/cm$^2$ in both the experiment and the calculation.
The coefficient $kn_{atom}$ was set to $8.55\times10^{16}$ m$^{-3}$.
$\gamma_s/2\pi$ was 127 Hz.
The measured contrast under continuous illumination was 2.4\% and the linewidth was 2.4 kHz.
The contrast decreased with increasing free evolution time owing to the coherence relaxation induced by the collision with other Cs atoms, the buffer gas, and the cell walls. 
The contrast exponentially decreased with increasing free evolution time from 0.2 ms to 1.1 ms.
The variation was 0.79\% per 1 ms.
The calculated resonance contrast had the same tendency as the measurement values.

The Q value as a function of the free evolution time is shown in Fig. \ref{fig:free_Q}.
The Q value linearly increased with increasing free evolution time because the linewidth is inversely proportional to the free evolution time $T$.
All the measured Q values were larger than the calculated values.
One of the reasons for this difference is that the experimental free evolution time is long because of the relationship between the atomic motion and the interaction region.
For example, the atoms falling in the dark state moved outside the interaction region during the first pulse, and they returned to the interaction region as a result of the collisions with buffer gas atoms after the free evolution time, where they were irradiated by the light again.
Because the effective free evolution time was long in the experiment, the measured Q values were larger than the calculated values.

The short-term frequency stability is determined as the product of the SNR and the Q value.
The contrast decreased with increasing free evolution time, and the Q value proportionally increased with increasing free evolution time.
Therefore, there is an optimal free evolution time $T$ for maximizing the product of the SNR and the Q value.
The optimal free evolution time was 1.0 ms in this experiment.
The maximum value was about five times larger than that under continuous illumination,
and it is expected that the short-term stability was improved by a factor of five.



\subsection{Contrast as a function of observation time}

\begin{figure}[t]
\includegraphics[width=\hsize,keepaspectratio=true]{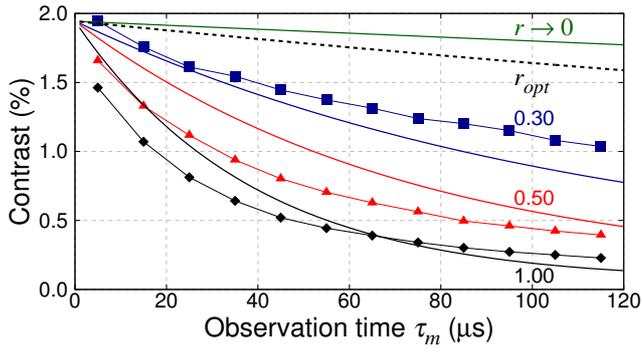}
\caption{(Color online) Contrast as a function of the observation time for different values of the observation intensity ratio $r$: the dashed line is a contrast with $r_{opt}$ that suppreses the variation of the light shift with the observation time{\cite{yano2015two}}.}
\label{fg:observation_contrast}
\end{figure}

Figure {\ref{fg:observation_contrast}} shows the contrast as a function of the observation time for different $r$.
The contrast had a maximum value at the pulse rise ($\tau_m \sim 1/\Gamma$) and exponentially decreased with increasing observation time.
At a very short observation time ($\tau_m < 1/\Gamma$), the contrast was small because of the limitation of the optical transition response.
When the observation time was larger than $1/\Gamma$, the contrast decreased with increasing observation time because of the repumping to the steady dark state{\cite{guerandel2007raman}}.
On the other hand, there was little change in the Q value with the observation time.
Therefore, it is necessary to reduce the observation time to obtain good short-term frequency stability.

In the conventional observation method ($r=1.0$), the contrast substantially decreased after a pulse rise.
A larger decrease in contrast was observed when a shorter observation time was set.
The decrease in contrast was 0.078 \%/$\mu$s at $\tau_m=$ 5.0 $\mu$s in this experiment.
This was two orders of magnitude larger than that with increasing free evolution time.
The contrast under the RR scheme was dominated by the observation time.

In the two-step pulse observation ($0.0<r<1.0$), higher contrast was obtained.
Because there was little repumping under a small $r$, a smaller decrease in contrast was observed for a small $r$.
When $r$ was negligible small and approached zero, the contrast increased and gradually approached a contrast curve with $r \sim 0$.
Therefore, the contrast curve with $r \sim 0$ is an asymptotic solution and is the upper limit of the contrast improvement in two-step pulse observation.
However, the contrast also tended to decrease when setting $r\sim 0$.
In such a case, the decrease in contrast is strongly dependent on the relaxation between the ground states $\gamma_s$, and it increased with increasing $\gamma_s$. 
Therefore, the decrease in contrast under $r\sim0$ was induced by the relaxation of the atomic state during the observation time.
The decrease was reduced to about half of that in the conventional method.
Therefore, the contrast increased with decreasing $r$, and the contrast in the two-step pulse observation ($r=0.3$) was twice that in the conventional method at $\tau_m = 25$ $\mu$s.
The decrease in contrast was lower for a larger observation time, 
and the contrast in the two-step pulse observation was 4.5 times larger than that in the conventional method at $\tau_m = 115$ $\mu$s.
These results show that a large contrast can be obtained by two-step pulse observation without reducing the observation time.

The light shift is dependent on the observation time $\tau_m$.
Our previous paper showed that the variation of light shift with observation time is suppressed by adjusting the observation intensity ratio $r${\cite{yano2015two}}.
In this calculation, the value of $r$ that suppresses the light shift variation was 0.04.
The contrast for this value of $r$ is lower than that for $r$ = 0.00; however, the maximum difference is no more than 10\% at $\tau_m$ = 120 $\mu$s.
Therefore, when the observation intensity ratio $r$ is adjusted to an appropriate value, both good contrast and low light shift variation can simultaneously be obtained.

\section{Conclusion}
We theoretically and experimentally discussed contrast improvement by two-step pulse observation.
From the results for the contrast and Q value as functions of the free evolution time, the contrast decreased with increasing free evolution time, whereas Q value proportionally increased with the free evolution time.
The results show that there is an optimal value at which the short-term frequency stability is maximized.
From the results for the contrast as a function of the observation time, the contrast significantly decreased with increasing observation time.
Also, the rate of decrease of the contrast with increasing observation time was two orders of magnitude larger than that with increasing free evolution time, meaning that the contrast under pulse excitation was strongly dependent on the observation time.
Two-step pulse observation substantially reduced the rate of decrease of the contrast with increasing observation time.
As a result, good contrast relative to that in the conventional method was obtained.
These results show that two-step pulse observation leads to improved short-term stability.
In addition, our previous paper showed that two-step pulse observation also reduces the light shift\cite{yano2015two}.
Two-step pulse observation is an effective method for realising CPT atomic clocks with good short-term and long-term frequency stability.

\section*{Appendix A: Numerical calculation}
\label{sec:ApdxA}
Here, the behavior of the dynamic density matrix is derived from the solution of Eq. (\ref{eq2}).
The easiest way to solve the time-dependent behavior of the density matrix is to perform numerical integration by the finite-difference method.
However, finite-difference methods require a long computation time because the time step must be smaller than the relaxation time $\Gamma^{-1}$ to satisfy the stability condition.
In this calculation, we show how to calculate the time dependent behavior of the density matrix using its eigenvector.
Since the numerical solution does not require numerical integration, it can reduce both the computation time and the error.

The total number of calculation elements is 9 because the density matrix is a $3 \times 3$ Hermitian matrix.
For the sake of simplicity of the differential equation, a vector comprising the nine elements $\vec{\rho}$ is defined as follows:

\begin{equation}
\vec{\rho}:=
\begin{pmatrix}
\rho_{11}\\\rho_{22}\\\rho_{33}\\{\rm Re}(\rho_{12})\\{\rm Re}(\rho_{13})\\{\rm Re}(\rho_{23})\\{\rm Im}(\rho_{12})\\{\rm Im}(\rho_{13})\\{\rm Im}(\rho_{23})
\end{pmatrix},
\end{equation}
Using $\vec{\rho}$, Eq. (\ref{eq2}) is rewritten as$^{\dagger}$
\begin{equation}
\frac{\partial \vec{\rho}}{\partial t}=\tilde{{\rm H}}\vec{\rho}.
\label{eq:eq7}
\end{equation}

\begin{table*}[t]
\normalsize

$^{\dagger}$
\begin{equation}
\tilde{{\rm H}}=
\begin{pmatrix}
-\gamma_s&\gamma_s 	&\Gamma_{31} &0&0&0&0 &\Omega_p&0\\
\gamma_s &-\gamma_s &\Gamma_{32} &0&0&0&0 &0&\Omega_c\\
0&0&-(\Gamma_{31}+\Gamma_{32})	 &0&0&0&0 &\Omega_p&\Omega_c\\
0&0&0&-\gamma_s&0&0&-(\delta_p-\delta_c)&\Omega_c/2&\Omega_p/2\\
0&0&0&0&-\gamma_f&0&\Omega_c/2&-\delta_p&0\\
0&0&0&0&0&-\gamma_f&-\Omega_p/2&0&-\delta_c\\
0&0&0&\delta_p-\delta_c&-\Omega_c/2&\Omega_p/2&-\gamma_s&0&0\\
-\Omega_p/2&0&\Omega_p/2&-\Omega_c/2&\delta_p&0&0&-\gamma_f&0\\
0&-\Omega_c/2&\Omega_c/2&-\Omega_p/2&0&\delta_c&0&0&-\gamma_f\\
\end{pmatrix}\nonumber
\end{equation}
\vspace{1mm}\hrule height 0.1mm depth 0.1mm width 173mm\vspace{2mm}
\end{table*}
In the Raman--Ramsey scheme, $\tilde{{\rm H}}$ during the excitation time is different from that during the free evolution time.
The matrix $\tilde{{\rm H}}$ is denoted as $\tilde{{\rm H}}_{\rm on}$ and $\tilde{{\rm H}}_{\rm off}$ when the pulse is on and off, respectively.
Also, the vector at the boundary (vector at pulse rise $\vec{\rho}_{\rm on}$ and fall $\vec{\rho}_{\rm off}$) is defined as follows:
\begin{equation}
 \begin{split}
  \vec{\rho}_{\rm on} &=\vec{\rho}(n(\tau+T)),\\
  \vec{\rho}_m &= \vec{\rho}(n(\tau+T)+\tau_m),\\
  \vec{\rho}_{\rm off} &=\vec{\rho}(n(\tau+T)+\tau),
  \end{split}
 \end{equation}
where $n$ is an integer.
Because $\vec{\rho}_{\rm on}$ is the vector $T$ s after that at the pulse fall $\vec{\rho}_{\rm off}$, and $\vec{\rho}_{\rm off}$ is the vector $\tau$ s after that at the pulse rise $\vec{\rho}_{\rm on}$,
the relationship between $\vec{\rho}_{\rm off}$ and $\vec{\rho}_{\rm on}$ can be expressed as follows:


 
 \begin{equation}
 \begin{split}
 \vec{\rho}_{\rm on} &= \exp(\tilde{{\rm H}}_{\rm off}T)\vec{\rho}_{\rm off},  \\
 \vec{\rho}_{m} &= \exp(\tilde{{\rm H}}_{m}\tau_m)\vec{\rho}_{\rm on},  \\
 \vec{\rho}_{\rm off} &= \exp(\tilde{{\rm H}}_{\rm on}(\tau-\tau_m))\vec{\rho}_{m},
  \end{split}
 \label{eq:on_off}
 \end{equation}
 
\noindent
where exp is the exponential function of the matrix.

By rearranging Eq. (\ref{eq:on_off}), the following equation can be obtained: 
\begin{equation}
\Bigl({\rm E}-\exp(\tilde{{\rm H}}_{m}\tau_m)\exp(\tilde{{\rm H}}_{\rm off}T)\exp(\tilde{{\rm H}}_{\rm on}(\tau-\tau_m))\Bigr)\vec{\rho}_m=0,
\label{eq:eigen}
\end{equation}
where E is the identity matrix.
From Eq. (\ref{eq:eigen}), because $\vec{\rho}_m$ is not a null vector, $\vec{\rho}_m$ is the eigenvector of the eigenvalue 0 of $\Bigl({\rm E}-\exp(\tilde{{\rm H}}_{m}\tau_m)\exp(\tilde{{\rm H}}_{\rm off}T)\exp(\tilde{{\rm H}}_{\rm on}(\tau-\tau_m))\Bigr)$. 
Since all the matrix elements are known, the vector at the pulse rise $\vec{\rho}_m$ can be derived by calculating the eigenvector of the matrix.
Note that the vector is normalized to satisfy the normalization condition of Eq. (\ref{eq3}).

\section*{Appendix B: Contrast derivation from density matrix}
Here, the relationship between the contrast and the density matrix of Eq. ({\ref{eq:contrast}}) is derived.

The photon numbers density $n_p$ is:
\begin{equation}
n_p=\frac{I}{\hbar \omega c},
\label{eq:photon_n}
\end{equation}
and the number density of absorbed photons per unit time $n_v$ is:
\begin{equation}
n_v=\frac{1}{3}n_{atom}\Omega {\rm Im}(\rho).
\label{eq:photon_v}
\end{equation}
From Eqs. ({\ref{eq:photon_n}}) and ({\ref{eq:photon_v}}), the time rate of absorbed photons is
\begin{equation}
\frac{n_v}{n_p}=\frac{\hbar \omega c n_{atom} \Omega {\rm Im}(\rho)}{3I}.
\end{equation}
The absorption index $\alpha$, which is the number rate of absorbed photons per unit length, is derived as:
\begin{equation}
\alpha	= \frac{n_v}{cn_p}=\frac{\hbar \omega n_{atom} \Omega {\rm Im}(\rho)}{3I}.
\label{eq:alpha}
\end{equation}
The transmitted light intensity $I_{tr}$ is:
\begin{equation}
I_{tr}= I \exp{(-\alpha L)}.
\end{equation}
Because $\alpha L$ is small, taking into account the fact that the conventional resonance contrast is no more than 10\%,
we can assume that
\begin{equation}
I_{tr} \approx I(1-\alpha L).
\label{eq:Itr}
\end{equation}
From Eqs. ({\ref{eq:alpha}}) and ({\ref{eq:Itr}}), the resonance amplitude $I_{pp}$ is:
\begin{equation}
I_{pp}=\frac{\hbar \omega n_{atom}L }{3}\Delta\Omega {\rm Im}(\rho).
\end{equation}
The contrast is defined as the signal amplitude divided by the background signal level.
The background level is proportional to $r$ and $k$ is a proportional constant that is determined by the sum of the laser intensities not interacting with alkali atoms such as higher-order sidebands.
\begin{equation}
{\rm Contast}(\%)=\frac{I_{pp}}{rI_p}\cdot 100=k\frac{\hbar \omega n_{atom}L }{3rI_p}\Delta\Omega {\rm Im}(\rho)\cdot 100
\end{equation}
Thus, the contrast can be derived using the density matrix.

\section*{Acknowledgements}
This work was supported by a Grant-in-Aid for JSPS Fellows (No. JP26$\cdot$6442).
The authors are grateful to Ricoh Company Ltd. for providing us with the Cs $D_1$ VCSEL.
The research of M. K. was supported by a Grant-in-Aid for Scientific Research (B) (Grant No. JP25287100), a Grant-in-Aid for Scientific Research (C) (Grant No. JP16K05500), and a Grant-in-Aid for Exploratory Research (Grant No.  JP 15K13545) from the Japan Society for the Promotion of Science (JSPS).
\bibliographystyle{spphys.bst}
\bibliography{Ref.bib}

\end{document}